\documentclass{article}

\def\half{{1 \over 2} } 
\def\Dslash{\slash \negthinspace \negthinspace \negthinspace \negthinspace D}
\def\ben{\begin{equation}}
\def\een{\end{equation}}
\def\bea{\begin{eqnarray}} 
\def\eea{\end{eqnarray}}
\input amssym.def
\input amssym.tex
\begin{document}

\hfuzz=100pt
\title{ The emergent  nature of time and the complex numbers
 in quantum cosmology}       
\author{G W Gibbons 
\\
D.A.M.T.P.,
\\ Cambridge University,
\\ Wilberforce Road,
\\ Cambridge CB3 0WA,
 \\ U.K.
}
\maketitle

\begin{abstract} 
The nature of time
in quantum mechanics is closely related to the
use of a complex, rather than say real, Hilbert space.
This becomes particularly clear when considering quantum field theory
in time dependent backgrounds, such as in cosmology, when
the notion of positive frequency ceases to be well defined.
In spacetimes lacking time orientation, i.e without the possibility
of defining an arrow of time, one is forced to abandon 
complex  quantum mechanics.
One also has to face this problem in quantum cosmology.
I use this to argue  that this suggests that, at a fundamental level,
quantum mechanics may be really real with not one, but
 a multitude of complex structures. I relate these ideas to other suggestions
that in quantum gravity time evolution  may not be unitary, possibly
implemented by a super-scattering matrix, and the status of CPT.
\end{abstract}

\section{Introduction}
The  topic of this conference  is {\it The Arrow of Time},
but before asking that we should
ask {\it What is the nature of time?}

Both Quantum Mechanics and General Relativity have something to say about this.

But what they say is not quite compatible  

For example, in  quantum mechanics,
there may be observables or operators
corresponding to  spatial  positions but time is 
not an  observable, i.e. it is not an operator
\cite{Wightman,UnruhWald,Giannitrapani}.
\footnote{see Pullin's contribution in this volume.}
More precisely, by an argument going back to Pauli,
commutation relations like 
\begin{equation}
\bigl[{\hat x} ^\mu,  {\hat P} _ \nu \ \bigr]= i \delta ^\mu _\nu
\end{equation}
are incompatible with the spectrum of ${\hat p}^\mu$
lying in the future lightcone.

In General Relativity on the other hand, 
space and time are usually held to be on the same footing.   

Because  the nature of time in Quantum Mechanics
is less familiar and less frequently  discussed,
than it is in General Relativity I shall begin by 
recalling \cite{Stueckelberg,Gibbons,GibbonsPohle,Gibbons1,Gibbons2,Gibbons3,Gibbons8}
        how 
{\sl time is intimately connected with the complex }{(Hilbert Space)} 
{ \it structure
of quantum  mechanics.}

In other words, the use of complex numbers and hence of  complex amplitudes
in Quantum Mechanics is intimately bound up with how 
Quantum States  evolve in time.  
\begin{equation} 
i {d \Psi \over dt} = H \Psi\,.
\end{equation}

In particular there can be no evolution if $\Psi$ is real
\footnote{Conversely, as shown by Dyson \cite{Dyson} 
in  his three-fold way, 
if $H$ is time-reversal invariant
one may pass to a real (boson)  or quaternionic (fermion)  basis}.

To proceed it is helpful to contemplate
more deeply than is usual in cosmology 

\section{The Structure of Quantum Mechanics}

If one analyzes the {\it Logical Structure
of Quantum Mechanics}  
one discovers that it consists of two different types of statements:
\begin{itemize}
\item {\bf I} Timeless\footnote{Of course in splitting the discussion
into two parts,  in  Part {\bf I} we take  the view that  Quantum
 Logic like its classical  Aristotelian special case is timeless.
This avoids appealing to Temporal Logic 
to resolving  such paradoxes as that of \lq \lq the  sea fight tomorrow
\rq \rq \cite{Cahn,Wright,Segre} and  puts the burden
of its resolution firmly where it belongs, in  Part {\bf II}. } 
  statements about {\it states, propositions, 
the Principle of Superposition, probabilities,  
observables}  etc  
\item {\bf II} Statements about how states  and observables change,
{\it  Schr\"odinger's equation and Unitarity}  etc. 
\end{itemize}

The upshot of an analysis of Part {\bf I} (so called Quantum Logic)
\cite{Jauch,JauchPiron}  
is that pure states  are
points in a {\it Projective Space} over ${\Bbb R},
{\Bbb C}\, {\rm or}\,{\Bbb  H}$ \footnote{By the principle of binary
  coding,
Classical Boolean  Logic may, for finite  sets at least,
be thought of as projective geometry over the Galois field of two
elements.  We  shall also ignore the exceptional case of the
octonions}.    
\begin{equation}
\Psi \equiv \lambda  \Psi \,,\qquad \lambda \in  
{\Bbb R},{\Bbb C}\, {\rm or}\,{\Bbb H}\,.  
\end{equation}

Now any vector space  over ${\Bbb R},{\Bbb C}\, {\rm or}\,{\Bbb  H}$ 
is a vector space $V$ over ${\Bbb R }$
{\sl with some additional structure}(cf. \cite{Stueckelberg,Myrheim}) , 
so let's use real notation. 
Observables  are {\sl symmetric}  bilinear forms:
\begin{equation}
\langle \Psi O \Psi \rangle  = \Psi^a O_{ab} \Psi ^b  \,,\qquad O_{ab}=O_{ba}
\,.\end{equation}
$a=1,2,\dots,n={\rm dim}_{\Bbb R}\, V$.
Mixed states $\rho $ are positive definite observables  dual
to the observables 
\begin{equation}
\langle  O  \rho  \rangle   = \rho ^{ab} O_{ab}  =  {\rm Tr} \, 
( \rho O )   \,, \qquad \rho^{ab}= \rho ^{ba} 
\,.\end{equation}
There is a privileged density matrix  
{\it the completely ignorant density matrix} which we may think
of as a {\sl  metric}\footnote{strictly speaking the inverse} 
 $g_{ab} $ on $V$ and use it
to normalize our states
\begin{equation}
\langle \Psi |\Psi  \rangle = g_{ab} \Psi ^a \Psi ^b\,,\qquad g_{ab}=g_{ba} 
\,, \qquad {\rm Tr} \rho = g^{ab} \rho_{ab} \,.\end{equation}

The upshot of a conventional   analysis of {\bf II} (Dirac called it 
               {\it Transformation Theory}) is that states change by
by means of linear maps
which preserves the metric (i.e. preserves complete ignorance) 
\begin{equation}
\Psi ^a \rightarrow S ^a \,_b \Psi ^b\,,\qquad g_{ab} S^a \,_c S^b\,_d =g_{cd}
\,.\end{equation} 
Thus $S \in SO(n,{\Bbb R})$, $n={\rm dim}_{\Bbb R} \,V $.
Infinitesimally
\begin{equation}
S^a\,_b = \delta ^a_b + T^a\,_b + \dots \,,
\end{equation}
where the  endomorphism or {\it Operator} \, $T^a\,_ b$ 
gives a {\it  two-form} when the index is lowered  
\begin{equation}
g_{ab} T^b\,_c := T^\flat  _{ac}= - T^\flat _{ca \,.} 
\end{equation}

But Dirac taught us that, just as in Hamiltonian mechanics,
{\it to every (Hermitian) Operator there is an Observable and 
{\it vice versa}.}
How can this be? Our vector space $V$ over ${\Bbb R}$
 needs some extra
structure, in fact a {\it complex structure}
$J^a \,_b $ or  {\it privileged operator}
 which also preserves the metric
(i.e. preserves complete  ignorance). 
\begin{equation}
 g_{ab} \,  J^a \, _ c J^b \, _ d = g_{cd} \,.  
\end{equation}
Then   \begin{equation}
J^a \,_b J^b\,_c=- \delta ^a_c \,\Longrightarrow 
 \,  \omega _{ab}=-\omega _{ba}\,,
\end{equation}
where the {\it symplectic two-form} 
$\omega_{ab}= g_{ac}  J^c \,_b $  may be used to lower indices
and obtain a symmetric tensor  for every (Hermitian) observable
(i.e. one that generates a transformation preserves the symplectic form)
 \begin{equation}
\omega _{ab} T^b\,_c := {T_\flat}   _{ac}= + {T_\flat}  _{ca} 
\,. \end{equation}

We can think of this more group theoretically \footnote{Or recall
what we might know about K\"ahler manifolds; 
Quantum Mechanics makes use of a 
K\"ahlerian vector space}.
{\sl In regular  Quantum Mechanics} 
 $V$ is a Hermitian vector
space its transformations should be unitary, but
\begin{equation}
U({n \over 2},{\Bbb C}) = SO(n,{\Bbb R}) \cap GL ({n \over 2} ,{\Bbb C}) 
\,,\end{equation}  
where $GL ({n \over 2} ,{\Bbb C} ) \subset GL(n,{\Bbb R})$ 
is the subgroup preserving $J$, and $ SO(n,{\Bbb R})  \subset GL(n,{\Bbb R})$
is the subgroup preserving the metric $g$.One also has 
\begin{equation}
U({n \over 2},{\Bbb C}) = SO(n,{\Bbb R}) \cap Sp(n, {\Bbb R})  
\,,\end{equation}
where $ Sp(n, {\Bbb R}) \subset GL(n,{\Bbb R})$ 
is the subgroup preserving the symplectic form $\omega $,
and of course 
\begin{equation}
U({n \over 2},{\Bbb C}) = Sp(n,{\Bbb R}) \cap   GL ({n \over 2} ,{\Bbb C})
\,.\end{equation}

\subsection{A precautionary principle}

Now the main message of this review  is that given a 
vector space $V$ over ${\Bbb R}$  
{\sl  it may have no complex structure} ($n$ must obviously be even!)
or if it is does, {\sl the complex structure 
may not be unique} (they are typically members of infinite families)

Thus on four dimensional Euclidean space ${\Bbb E}^4$ 
they belong (modulo a choice of orientation) to a  two-sphere  
$S^2 = SO(4)/U(2)$.

More generally, every  quaternion vector space has such a   2-sphere's
worth of complex structures \footnote{cf Hyper-K\"ahler manifolds such as K3}
, i.e. {\sl a  2-sphere's worth of
of times}!

To bring out the fact that in physics we use many different complex
structures for many different reasons it is occasionally
helpful to indicate explicitly by the symbol $i_{\rm qm}$
the very particular complex structure on the  Hilbert
space ${\cal H}_{\rm qm}$  of the standard model and so that
Schr\"odinger's equation really reads  
 
\begin{equation} 
i_{\rm qm}  {d \Psi \over dt} = H \Psi\,.
\end{equation}

At a  more mundane level, the use of the notation $i_{\rm qm}$     
brings out  how dangerous
and misleading, certainly to the beginner,  
it can be to use complex notation too sloppily. Suppose one has
a theory, with an $SO(2)$ symmetry (gauged or un-gauged) . It is tempting 
to collect the fields, e.g. scalars $\phi_1,\phi_2$   in pairs
\begin{equation}
\phi=\phi_1+i \phi_2\,. \label{trap} 
\end{equation}
 Now the $i$, which generates the $SO(2)$ 
 action  is (\ref{trap}) is not the same  as $i_{\rm qm} $.
This is clear from the fact that charge conjugation
 \begin{equation}
C:  \phi_1+i \phi_ 2\rightarrow \phi_1-i \phi_2
\end{equation}
is anti-linear, i.e anti commutes with $i$ but is nevertheless
represented on  ${\cal H}_{\rm qm}$ as a linear operator,
i.e. one which commutes with  $i_{\rm qm} $ .
Note that there would be no temptation to indulge in 
such notational confusion if there were three scalar fields 
$\phi_1, \phi_2,\phi_3$ and the  symmetry $SO(3)$.

Just how confusing the sloppy use of 
the somewhat ambiguous  complex notations currently  can be is 
nicely illustrated in \cite{Saunders}
in the context of quantum field theory, an example which will be of relevance
later.

At a purely practical level, the avoidance of
an excessive use of  complex notation also helps in  
formulating action  principles in an intelligible fashion.  
One is often instructed that in varying an action with, for example
complex scalars, that one should vary the action regarding
\lq $\phi$ and its complex conjugate $\bar \phi$ as independent\rq.
On the face of it this sounds ridiculous. What is actually meant is
that one varies regarding the real $\phi_1$  and imaginary  $\phi_2$ parts
of $\phi$ as independent. It is easy to check that,
 as long as the action is real, then  this cook book
recipe will give the correct result, essentially because
varying with respect to $\bar \phi$  gives the complex conjugate
of the equation obtained by varying with respect to $\phi$.
However as a general principle the cook book recipe cannot be
of general validity. It fails, and is inconsistent,  if,  for example,
one varies a complex valued function of a complex variable and its
complex conjugate. If it only works in special cases
and can lead to incorrect results, it seems best to avoid both the cook book
recipe 
and the misleading notation that gives rise to it.

In conclusion therefore, 
it seems wise  to adopt a course  of action, particularly
at the classical level before quantization, in which 
one proceeds as far as possible by considering all physical quantities
and their related mathematical structures to be real until 
one is forced to introduce complex notation and $i_{\rm qm}$ 
at the point where one introduces quantum mechanics.   

In other words, in what follows,  I plan to follow, in so far as is possible,
 Hamilton's  course of action \cite{Hamilton}

\begin{quote}
The author acknowledges with pleasure that
he agrees with M. CAUCHY, 
in considering every (so-called) Imaginary Equation as a symbolic 
representation of two separate Real Equations: but he differs from
that excellent mathematician in his method generally, and especially
in not introducing the sign
 $\sqrt{-1}$ until he has provided for it,
 by his Theory of Couples, a possible and real meaning,  as a symbol
of the couple $(0,1)$. 
\end{quote}

\subsection{Dyson's Three-fold way}
In this language, Dyson's observation \cite{Dyson}  
is that in standard quantum mechanics an anti-linear involution
$\Theta$ acting on rays may be normalized to satisfy
\begin{equation}
\Theta ^2 =\pm 1\,,
\end{equation}
where the plus sign corresponds to an even spin state and
the odd sign to an odd  spin state.To say that $\Theta$
is anti-linear is to say that it anti-commutes with the 
standard  complex structure $i_{\rm qm} $, 
${ i_ {\rm qm} }  ^2=-1$

\begin{equation}
\Theta i_{\rm qm } + i_{\rm qm } \Theta =0\,.
\end{equation}
Now for the plus sign $\Theta$, is a projection operator
and we get  what is called a {\it real structure} 
on the original complex Hilbert space and if the Hamiltonian
is time-reversal invariant, then we may use the projection
operator to project onto the subspace of real states.
On the other hand for the minus sign
we construct

\ben
        K = \Theta \, i_{\rm qm}  
\een

and find that $\Theta, i  _{\rm qm} , K$ satisfy
the algebra of the quaternions.

\subsection{Relation to Jordan Algebras} 
There is an interesting tie in here with the theory
of Jordan algebras \cite{Townsend1,Gibbons6}  which were  originally 
introduced by Jordan as a possible avenue for
generalizing quantum mechanics but in the end
led to the same three basic possibilities.  

In all three varieties of quantum mechanics the 
states, i.e. the space of positive semi-definite  Hermitian
matrices in ${\rm Herm} _n({\Bbb K})$, ${\Bbb K} = {\Bbb R}, {\Bbb C},
 {\Bbb H}$
form a homogeneous convex self-dual cone,
Moreover as observed by Jordan, 
they satisfy an abelian, but non-associative,  algebra
whose multiplication law is
\begin{equation}
\bigl ( O_1, O_2 \bigr )
 \rightarrow {1 \over 2}  \bigl ( O_1 O_2 + O_2 O_1 \bigr  )\,.
\end{equation}
The algebra $J^{\Bbb K}_n$     thus obtained  
is real and power law associative and
thus belongs to the class of what are now known as  {\it Jordan Algebras}.  

In fact the  list of finite dimensional irreducible 
homogeneous self-dual cones is quite small and coincides with
the list of finite dimensional irreducible Jordan algebras. The list is:

{\center
\vskip.5cm
\begin{tabular}{|c|c|c|c|}\hline
Cone &   Algebra &  Reduced Structure Group & Automorphism Group  \\ 
\hline\hline
$C({\Bbb E} ^{n-1,1})$ & $\Gamma(n-1)$  &  $SO(n-1,1)$ & $SO(n-1)$\\
\hline
$C_n({\Bbb R} )$ & $J_k^{\Bbb R} $ & $ PSL(k;{\Bbb R}  )$  & $SO(n)$  \\
\hline
$C_n({\Bbb C} ) $ & $J_j^{\Bbb C} $ & $PSL(n;{\Bbb C} )$ & $SU(n)$  \\
\hline
$C_n({\Bbb H} ) $ &$J_k^{\Bbb H} $  & $SU^\star (2n) $ &  $Sp(n)$ \\  
\hline
$C_3({\Bbb O } )$ & $   J_3^{\Bbb O}  $ & $E_{6(-26)}$ &$F_4$\\
\hline
\end {tabular}
\vskip .5cm
}

\medskip \indent \indent 
{$\bullet$} $C({\Bbb E}  ^{n-1,1})  \subset \Gamma(n-1)$
 is the usual Minkowski cone,
in ${\Bbb E}^{n-1,1}$ based on a the sphere $S^{n-2}$.
The automorphism group is the  Lorentz group
$SO(n-1,1)$.

\medskip \indent \indent {$\bullet$} $C_k({\Bbb R}) \subset
J^{\Bbb R}_n$: the set of positive semi-definite
$n\times n$ real symmetric
matrices. The reduced structure groups is $PSL(n,{\Bbb R})$
and the automorphism group is $SO(n-1)$.

\medskip \indent \indent  {$\bullet$} $C_n({\Bbb C}) \subset
J^{\Bbb C}_n$: the set of positive semi-definite
$n\times n$ hermitian
matrices. The reduced structure group  is $PSL(k,{\Bbb C})$
and the automorphism group is $SO(n)$.

 \medskip \indent \indent  {$\bullet$} $C_n({\Bbb H}) \subset J^{\Bbb H}_n$:
$n\times n$ positive definite quaternionic hermitian
matrices. The reduced structure group is $SU^\star (2k)$
and the automorphism group is $Sp(k)$.

 \medskip \indent \indent  {$\bullet$} $C_3({\Bbb O}) \subset
J^{\Bbb O}_3$: the set of positive semi-definite
$3\times 3$ octonionic hermitian
matrices. reduced structure group is  is $E_{6(-26)}$
and the automorphism group is $F_4$.

In all cases the  automorphism group $Aut(J)$
of the Jordan algebra $J$ is the stability group
of the unit element in the algebra, which may be taken as a unit matrix.
The reduced structure group of the algebra $St_0(J)=PLSG$ 
is the subgroup of the structure group $G=Str(J)$,
leaving the norm of the Jordan algebra invariant.

Note that these results subsume the foundational
Alexandrow-Zeeman \cite{A,Z} 
theorem which states that the auto-morphism group
 of the causal structure of Minkowski spacetime
(defined by the cone $C({\Bbb E} ^{n-1,1}) $ ) consists of dilations
and Lorentz transformations \cite{H}.

\medskip \noindent In the case of $\Gamma(n-1)$ one
may think of $v$ as an element of the Clifford algebra
${\rm Cliff}(n-1,1;{\Bbb R})$, on sets $v=v^\mu \gamma_\mu$.
However the Jordan algebra $\Gamma(n-1)$ 
is generated by $\gamma_i$ and the identity matrix. 
Then in all cases the commutative but not associative
Jordan product is given by one half the anti-commutator,
$u\bullet v={ 1\over 2} (uv+vu)$. 
The cone $C(J)$ is then obtained
by taking the exponential $\exp(v)$ of elements $v \in J$.
This is well defined because of the power associativity
property $v \bullet v^r=v^{r+1}$ of the algebra.

\subsection{Special Cases: low order isomorphisms}

It is a striking fact that in the case of $2 \times 2$ matrices
over ${\Bbb K}={\Bbb R},{\Bbb C}, {\Bbb H}$ 
the Jordan algebras coincide with the Clifford algebras,
fact perhaps more familiar in the form 
\begin{eqnarray}
Spin(2,1) &\equiv& SL(2,{\Bbb R}) \,,\\
Spin(3,1) &\equiv& SL(2,{\Bbb C})\,,\\
Spin(5,1) &\equiv& SL(2,{\Bbb H})\,.\label{isomorphism}
\end{eqnarray}
There is also a closely related statement over the octonions
for $Spin(9,1)$ which crops up in string theory.

The middle isomorphism in (\ref{isomorphism})
has lead Penrose to attempt, in his Twistor theory, to connect
the use of the complex numbers in spinor analysis
with that in quantum mechanics. A connection which moreover
seems  to  give a privileged position to four
spacetime dimensions. I think one can take a very different view
\cite{Gibbons5}
but to   appreciate it we need to make an excursion
into  

\section{Spacetime Signature and the Real Numbers}

The basic point being made here is that
in $4+1$, and indeed $9+1$ and $10+1$, spacetime dimensions,
it is possible, by choosing the spacetime signature appropriately,
to  develop spinor analysis {\sl at the classical level}
{\sl entirely over the reals}. That is, to consistently use
Majorana spinors whose components really are real. 
In four spacetime dimensions this requires the mainly plus
signature convention (the opposite to that which Penrose
uses). The complex numbers need only enter when one quantizes.

To see this in more detail we need some facts about
\subsection{Clifford Algebras} 

Given a vector space $V$ \footnote{$V$ is {\sl not} ${\cal H} _{\rm {qm}}$ 
thought of as real! A good reference for the properties of Clifford
algebras used here 
 is \cite{Dabrowski}, see also \cite{Ferrara} }  with metric $g$, of signature $(s,t)$ 
where $s$ counts the positive and $t$ the negative signs, 
 Clifford algebra ${\rm Cliff}(s,t; {\Bbb R})$ is 
by definition the  associative algebra over the {\sl reals}   generated
by the relations   
\begin{equation}
\gamma _\mu \gamma _\nu + \gamma _\nu  \gamma _\mu  = 2 g_{\mu \nu}\,,
\end{equation}
where $\gamma$ is a basis for $V$ .  
As a real algebra, the signature does make a difference.
For example
\begin{equation}
{\rm Cliff}(0,1;{\Bbb R}) \equiv {\Bbb  C}\,,
\end{equation} 
while
\begin{equation}
{\rm Cliff}(0,1;{\Bbb R}) \equiv {\Bbb  R} \oplus {\Bbb  R}\,.
\label{double}
\end{equation}
In  fact   $ {\rm Cliff}(0,1;{\Bbb R})$ is identical with   
what are often called  \lq double numbers \rq or \lq hyperbolic
numbers\rq , i.e
numbers of the form. 
\begin{equation}
a+eb\,,\qquad a,b \in {\Bbb R}\,,\qquad e^2 =1\,.
\end{equation} 
             
As an algebra, $ {\rm Cliff}(0,1;{\Bbb R})$ is not simple,
 $P_\pm= {1 \over 2}  (1\pm e )$ are projectors onto two commuting
sub-algebras.   

In a matrix representation 
 \begin{equation}
i= \pmatrix{ 0 &-1\cr 1 & 0 \cr} \,,\qquad e = \pmatrix{ 0 & 1\cr 1 & 0 \cr}\,.
\end{equation} 

However  if we pass to the complex Clifford over ${\Bbb C} $  we   
lose the distinction since
\begin{equation}
{\rm Cliff}(0,1;{\Bbb C}) \equiv {\rm Cliff}(0,1;{\Bbb C})\,
\equiv M_2({\Bbb C})\,,
\end{equation} 
where $ M_2({\Bbb C}) $ is the algebra of all complex valued
two by two matrices.

It is precisely at this point that the precautionary principle
comes in. We should not rush  into  adopting   
\begin{equation}
{\rm Cliff}(3,1;{\Bbb C}) \equiv {\rm Cliff}(1,3;{\Bbb C})
\equiv M_4({\Bbb C})\,,
\end{equation}
but rather enquire what are the possible differences 
between the two signatures \footnote{A similar point has been made recently
by Schucking \cite{Schucking} but he  plumps for the
quaternions} . 
In fact
\begin{equation}
{\rm Cliff}(3,1;{\Bbb R}) 
\equiv M_4( {\Bbb R})\,, 
\qquad{\rm Cliff}(1,3;{\Bbb R}) \equiv M_2({\Bbb H})   
\,,\label{difference} \end{equation}
where
\begin{equation}
{\Bbb H} \equiv {\rm Cliff}(0,2;{\Bbb R})
\end{equation}
are the quaternions.
Despite the differences the spin groups are identical 
\begin{equation}
{\rm Spin} (3,1) \equiv {\rm Spin}(1,3) \equiv SL(2,{\Bbb C})\,,
\end{equation}
but if discrete symmetries  are taken into account they differ:
\begin{equation}
{\rm Pin} (3,1) \ne  {\rm  Pin} (1,3)\,.
\end{equation}
This has important consequences  in spacetimes which are 
time, space  or spacetime  non-orientable
 \cite{Rohm,Carlip,Friedman,GibbonsChamblin,Chamblin}.

\subsection{Chiral rotations}

Independently of signature
\begin{equation}
\gamma _5= \gamma_0 \gamma_1 \gamma_2 \gamma_3 \,, \qquad  \gamma_5^2 =-1\,.
\end{equation}

Moreover if $\gamma_\mu$ generate a Clifford algebra, then so do
\begin{equation}
e^{\alpha \gamma_5} \gamma_\mu e^{-\alpha \gamma_5}= \cos 2 \alpha \gamma_\mu
+ \sin 2 \alpha \gamma _5 \gamma_\mu\,, \qquad \alpha \in {\Bbb R}\,. 
\end{equation}

Thus by choosing  $\alpha = \pi$ we can reverse the sign of the $\gamma_\mu$
and so we expect that no physical consequences should  follow
from the choice of sign. 
 
The chiral rotations maintain the reality properties  of the
gamma matrices. Multiplication by $i$ of course
reverses   the signature.

\subsection{Majorana Spinors}
It is a striking and, I believe, a possibly rather significant  fact that  
the signature $(3,1)$ leads directly to  a  Majorana representation,
in which all $\gamma$  matrices are real. Certainly 
if one holds  that $N=1$ supersymmetry and $N=1$ supergravity  are important,
this fact renders the mainly positive signature rather attractive. 
The precautionary principle would lead one to adopt
the signature $(3,1)$ and use a real notation for as long
as one can, certainly at the classical level where one need
never introduce complex numbers.
Thus the basic entities are Majorana spinors $\psi$ 
belonging to a four dimensional real vector space ${\Bbb M}$  with 
real, or real Grassmann number components $\psi ^a$, $a=1,2,3,4$. 

Note that if $\psi$ is a Majorana spinor then so is
its chiral rotation $e^{\alpha \gamma_5} \psi$. 

The charge conjugation matrix $C=-C^t $ satisfies
\begin{equation}
C \gamma _\mu C^{-1}= -\gamma _\mu ^t \,,
\qquad C\gamma _5 C^{-1}=- \gamma _5^t\,.
\end{equation}
It serves as a Lorentz-invariant symplectic form on ${\Bbb M}$.
Thus ${\rm Spin}(3,1) \subset {\rm Sp} (4;{\Bbb R}) \equiv {\rm Spin} (3,2)$.

\subsection{Dirac Spinors}

To incorporate Dirac spinors, one considers pairs of
Majorana spinors $\psi ^i$ , $i=1,2$ which are elements  of  
${\Bbb R} ^4 \oplus {\Bbb R} ^4 \equiv 
{\Bbb R} ^4 \otimes {\Bbb C}^2  \equiv {\Bbb R} ^8 $ 
If $\delta_{ij}$ is the metric and  $\epsilon_{ij}= \delta _{ik} J^k\,_j$,
the symplectic and $J^k\,_j$ the complex structure 
which rotates the two summands into each other, we can endow
${\Bbb D}\equiv {\Bbb R} ^8 $ with a symplectic form $\omega$   and a 
pseudo-riemannian metric $g$ , and hence
a pseudo-hermitean structure.
In components, for commuting spinors,
\begin{equation}
g(X,Y)  = X ^{ia} C_{ab} \epsilon_{ij} Y ^ {ja} =g(Y,X) 
\end{equation}
\begin{equation}
\omega (X,Y) =  X ^{ia} C_{ab}\delta _{ij} Y^ {ja}=  -\omega (Y,X)\,,
\end{equation}
so that
\begin{equation}
\omega(X,Y)= g(JX,Y)\,.
\end{equation}
The signature of the metric $g$ is $(4,4)$ 
and of the hermitian form, which is usually written
\begin{equation}
{\overline  \psi } \psi , 
\end{equation}
 where the Dirac conjugate 
\begin{equation}
{\overline \psi}= \psi ^\dagger \beta 
\end{equation} 
is $(2,2)$.
The \lq light cone\rq  on which 
${\overline  \psi } \psi $ consists of Majorana Spinors

Not that electromagnetic rotations and chiral rotations
commute with one another.

Alternatively we can think of the 
Dirac spinors as elements of a four dimensional complex vector space
${\Bbb D} = {\Bbb M} _{\Bbb  C}\equiv {\Bbb C} ^4$, 
the complexification  of the real space of of Majorana spinors ${\Bbb M}$.

\subsection{Weyl Spinors}
To see where Weyl spinors fit in we  observe that $\gamma_5$ 
acts as a complex structure converting 
${\Bbb M} \equiv {\Bbb R} ^4$ to $ {\Bbb W} \equiv {\Bbb C} ^2$. 
In other words,  we write 
\begin{equation}
{\Bbb M}   \otimes _{\Bbb R} {\Bbb C} = 
{\Bbb D } = {\Bbb W}   \oplus 
{\overline {\Bbb W  }  }\,, \label{summands} 
\end{equation}    

Elements of ${W } ^2$ are  chiral  spinors for which
\begin{equation}
\gamma _5 \psi_R  = i \psi_R,
\end{equation}
Elements of ${\overline {W }} $ are  anti-chiral  spinors for which
\begin{equation}
\gamma _5 \psi _L = -i \psi_L,
\end{equation}
The projectors ${1 \over 2}  (1 -i\gamma_5)$ and 
${1 \over 2}  (1+i\gamma_5)$ project onto chiral and anti-chiral
Weyl spinors respectively.

It is of course possible to treat  Weyl spinors without
the explicit introduction of complex numbers
at the expense of introducing pairs of  Majorana spinors
$\psi_1, \psi _2$ subject to the constraint that
\begin{equation}
\gamma_5 \psi _1 = -\psi _2\,,\qquad \gamma_5 \psi _2 = \psi _1\,.
\end{equation} 
One then has
\begin{equation}
\psi_R = \psi _1 +i\psi _2\,,\psi _L = \psi _1-i\psi _2 \,. 
\end{equation}

\subsection{Signature reversal non-invariance}

Many people would argue that after all, a choice of
signature is only a convention. That is true, but as we have seen
above, this  choice of convention comes with consequences.
Moreover reversal of signature is not a symmetry of
the basic equations of physics.
as  has emerged very clearly recently  in work 
aimed at understanding why 
the observed cosmological constant is so small
in comparison with its expected value. There have been a number
of suggestions \cite{tHooft,Kaplan,Erdem1}  
that this might be due to a symmetry, analogous to chiral symmetry
which is used to account for the smallness  of  of the electron mass.
One candidate for such a symmetry , which may be expressed in a manifestly
generally  covariant, and simple  fashion     
is the symmetry under change of spacetime signature
\ben
g_{\mu \nu} \rightarrow -g_{\mu \nu}\,. \label{flip} 
\een     .

For flat spacetime  this is equivalent to the transformation \cite{tHooft}
\ben
x^\mu \rightarrow ix^\mu \,,
\een
taking West  Coast to East  Coast, 
\ben
{\Bbb E} ^{3,1} \rightarrow {\Bbb E} ^{1,3}\,, 
\een
but complexifying or analytic continuation of coordinates 
are  not without problems in curved spacetime  and so I prefer 
(\ref{flip}) which does the job just as well. I have a similar prejudice
against formulations in terms of non-generally covariant concepts such as  
energy \cite{Kaplan}. 

Under (\ref{flip}) one has
\ben
R_{\mu \nu} \rightarrow R_{\mu \nu} 
\een
and so, if $\Lambda \ne 0$, 
(\ref{flip})    is definitely not a symmetry of the equations
\ben
R_{\mu \nu}= \Lambda g_{\mu \nu} \,.
\een 
and hence is violated by a non-vanishing cosmological constant.

If scalar fields are present, then  (\ref{flip}) is  violated
by mass or potential terms    
since under (\ref{flip}) the Christoffel symbols and hence the connection
are unchanged  
\ben
\{_\mu \,^\nu \,_\sigma  \} \rightarrow  
\{_\mu \,^\nu \,_\sigma  \} \,,\qquad \nabla _\mu \rightarrow \nabla _\mu \,, 
\een
but the equation 
\ben
g^{\mu \nu} \nabla _\mu  \nabla  _\nu \phi = V^\prime (\phi)   
\een
is not invariant and neither is the Einstein equation
\ben
{ 1 \over 8 \pi G} R_{\mu \nu} = 
\partial _\mu \phi \partial _\nu \phi + g_{\mu \nu} V(\phi)\,,  
\een

On the other hand, the source free   Maxwell equations are
invariant, but the Einstein equation.  
\ben
 { 1 \over 8 \pi G} R_{\mu \nu}= g^{\sigma \tau} 
 F_{ \mu \sigma} F_{\mu \tau } -{1 \over 4} g_{\mu \nu}
g^{\alpha \beta } g^{\sigma  \tau}  F_{\alpha \sigma } F_{\beta  \tau}    
\een
is not.

This is part of a more general pattern, 
for a massless $p$-form field strength  in $n$ spacetime dimensions 
(so that $p=1$ corresponds
to a scalar and in four dimensions, $p=3$ to a pseudoscalar or axion) 
then while the equation  of motion is invariant,  (\ref{flip}) induces 
\ben
T_{\mu \nu} \rightarrow (- 1)^{p+1} T_{\mu \nu} \,.
\een

From this it is clear that the Maxwell equations coupled to 
a complex scalar field, that is the Abelian Higgs or Landau Ginzburg  model
are not invariant. This can be seen from
\ben
\nabla _\nu  F^{\mu \nu}= J^\mu\,.
\een 
Under (\ref{flip}) 
\ben
F^{\mu \nu} \rightarrow   F^{\mu \nu} \,,
\een
but
\ben
J^\mu \rightarrow -J^\mu \,.
\een
Similarly the Lorentz equation 
\ben
{ d^2 x^\mu \over d \tau^2 } + 
\{_\sigma  \,^\mu \,_\tau   \} { d x ^\sigma \over d \tau} 
{ d x ^\tau \over d \tau} 
= { e \over m} g^{\mu \alpha} F_ {\alpha \beta}  
 { d x ^\beta  \over d \tau}\,,
\een
is not invariant under (\ref{flip}).

Thus the  classical  equations of motion of the  bosonic sector of the 
standard model are  certainly not
invariant under $(\ref{flip})$. To make them so would entail adding
additional fields whose energy momentum tensor is opposite in sign 
to the standard case. These fields would antigravitate  rather than gravitate. 
Various schemes of this sort have been discussed in the literature
(e.g. \cite{Linde1,Linde2,Linde3}).

If, therefore, the signature reversal is not a symmetry of our world,
then it seems  reasonable to me to suppose that one signature is preferred
over the other, and that is the view being advocated here. 

Of course one could follow Duff and Kalkkinen 
 that we have simply mistaken the
dimension we are in,  \cite{DuffKalkkinen1,DuffKalkkinen2}   
or conclude that the signature of spacetime may
vary from place to place, some regions having  signature
$-+++$ and some signature $+---$ \cite{Gibbons7}. 
Perhaps one should say that spacetime signature is an emergent property.  

\section{More than one time: Signature Change}
We have been arguing that time, or at least a 
universal  complex structure on the quantum mechanical  Hilbert space
single may be an emergent, or historical
phenomenon.  

This seems clearest  in certain, instanton  
based, approximate, treatments  of  the birth   
of the universe based on what Hartle and I have called 
Tunnelling Metrics \cite{GibbonsHartle} ,in  which a Riemannian
manifold $M_R$ and and a Lorentzian manifold $M_L$ are joined
on across a surface $\Sigma$ of time symmetry which may be regarded
as the origin of time surface. There is no time in $M_R$ where
the metric signature is $++++$. The metric signature flips to
$-+++$  across $\Sigma$. If that can happen why can't it flip
to $--++$ across some other surface, as  suggested by Eddington
long ago \cite{Eddington} ?  

Signature flip also arises in brane-world scenarios
in which the brane bends over in time   while
remaining a smooth sub-manifold of the Lorentzian bulk spacetime,ceases to be 
{\it timelike}, but rather spacelike with positive definite
(i.e. Riemannian) induced metric \cite{GibbonsIshibashi, Mars1,Mars2}.
However unless the bulk as more than one time the transition can only be from
Lorentzian to Riemannian. In the model studied in \cite{GibbonsIshibashi}
time certainly emerges after the collision of two branes.

The question therefore arises, could  two, or possibly more 
than two times have  emerged?
There has been a fairly large amount of work
on the possibility of two or more times, i.e
on spacetimes of signature $--,+,+,+\, $ or $-,-,-,+,+,+$.
A very early example is hinted at by Halsted \cite{Halsted} 
A  later, and for me 
difficult to understand example is  \cite{Bennett}, 
where the extra temporal
coordinate is called \lq anti-time \lq or \lq eternity \rq). 

As far as a can see,  little attempt to relate them to
the algebraic structure of quantum mechanics,
although a theory of Kostant comes quite close.

In fact, the standard reason for rejecting such theories
is  the  existence of the  instabilities and 
causality violations that result as a 
consequence of the fact that the interior of the light cone is no 
longer convex.
This is clearly shown by Dorling's argument \cite{Dorling} that
the lowest mass particle is such a spacetime could decay into 
particles of heavier mass.
 In Kaluza-Klein theories, timelike extra dimensions lead to
negative energies for vector fields on dimensional reduction
\cite{GibbonsRasheed} and provides limits on their size
\cite{Yndurain}.

One way to say this is that 
necessarily  such spacetimes cannot
admit a {\sl time orientation} and hence, in accordance with our
general outlook,  cannot admit standard complex  quantum mechanics.

Among multi-time theories, a 
particularly intriguing  case from the mathematical point of view
is that of six-dimensional manifolds with neutral or Kleinian
signature $(+++---$ . In other words where there is a
complete symmetry between space and time. This 
has been energetically pursued by Cole over many years
\cite{Cole1,Cole2,Cole3,
Cole4,Cole5,Cole6,Cole7,Cole8,Cole9,Cole10,Cole11} in an attempt
to make physical sense of it. I am skeptical but believe
it may ultimately play a role in string theory. 

One has the isomorphisms
\ben
SO(3,3) \equiv   SL(4,{\Bbb R})/{\Bbb Z}_2 \,,\qquad 
{\rm Cliff} (3,3;{\Bbb R})  \equiv 
M_8({\Bbb R})\,. 
\een
The first  isomorphism  links us to real three-dimensional projective
geometry, via a a real form of Twistor theory \cite{Gibbons5}.
One may think of ${\Bbb E}^{3,3}$ as the space of bi-vectors
in $L^{\mu \nu}  =-L^{\nu \mu}$ in  ${\Bbb R} ^4$ endowed with the metric  
\ben
{1\over 4} \epsilon_{\mu \nu \sigma \tau} L^{\mu \nu} L^{\sigma \tau}  \,.
\een
By the well-known Pl\"ucker correspondence,  
lines in ${\Bbb R} {\Bbb P} ^3$ correspond to 
simple bi-vectors in      and hence to to null 6-vectors in  ${\Bbb E}^{3,3}$.
It is also possible to regard ${\Bbb R} ^4$ or its projectivization
 ${\Bbb R} {\Bbb P} ^3$ as the space of Majorana spinors in four spacetime
dimensions. Conformally $SO(3,3)$ is the conformal
group of ${\Bbb E} ^{2,2} $. In Penroses's Twistor
Theory one complexifies and another real form is $SO(4,2)$ the conformal
group of ordinary Minkowski spacetime  ${\Bbb E} ^{3,1}$.

 Kostant \cite{Kostant,GuilleminSternberg}  has made the imaginative  proposal
that 
our spacetime (with signature $(3,1)$)  is
a 3-brane embedded in a  six-dimensional bulk spacetime
with a metric of signature $(3,3)$. The restriction
of the ambient metric to the normal bundle has
signature $(0,2)$ and the associated  $SO(2)$ symmetry allows
him to think of the normal bundle as a complex line bundle
over spacetime. This is the origin of electromagnetism in his theory.

To obtain examples, Kostant   noted that the conformal group
of  ${\Bbb E}^{3,3}$ is $SO(4,4)$. More accurately
$SO(4,4)$ acts globally on the conformal compactification    
of  ${\Bbb E}^{3,3}$, which may be regarded as the space of null rays
in ${\Bbb E}^{4,4}$. To get compactified Minkowski spacetime
($S^1 \times S^3/{\Bbb Z}_2 $ 
one intersects the null cone of the origin of  ${\Bbb E}^{4,4}$ with a 
6-plane through the origin of signature $(4,2)$. 
An interesting aspect   is that since we are dealing with $SO(4,4)$
there is a triality which acts. Mathematically, Kostant's model
has many intriguing  features (see \cite {Biedrzycki1,Biedrzycki2})
but so far clearly fails to make much contact with  the real world.

Another, purely technical use of three times 
is to study integrability.  
Because ${\Bbb E} ^{3,3} $ admits a para-hermitean structure, in
other words it admits an isometric involution $J$ on the tangent space
which , $J^2=1$,(i.e. para-complex) and such that $g(JX,JY) = g(X,Y)$,
this  may be used to obtain the KP equations
via a self-duality condition on $\frak{sl}(2,{\Bbb R})$ gauge fields
\cite{DasSezginKhavienga}.

\section{Examples}

The general algebraic 
considerations  considerations may seem rather abstract, but
they have already arisen  in the application of quantum mechanics
to cosmology. 

In what follows, I shall  give some examples.
Before doing so I note that 

\vskip .5cm 
{\sl The much discussed question of whether black hole evaporation
is unitary is {\bf meaningless} if there is no complex structure,
and ill-posed if there is more than one.} 
\vskip .5cm

\subsection{Quantum Field Theory in Curved Spacetime}
In  Quantum Field Theory in Curved Spacetime
the main problem is that there is no unique definition
of  \lq \lq  positive frequency \rq \rq  . In the free theory,  
 $V = {\cal H}_{\rm one \, particle} $ is the space of 
real-valued solutions of  wave equations.  
$V$ is   naturally (and covariantly)  a symplectic (boson), 
or orthogonal (fermion)\footnote{we use real (Majorana) commuting
spinors for convenience: there use is not essential} vector space
\begin{equation}
\omega(f,g) = \int \bigl (\dot f g - \dot g g \bigr ) d^3\,x
= -\omega(g,f)
\end{equation} 
\begin{equation}
g(\psi,\chi)  = \int \bigl (\psi ^t  \chi ) d^3\,x
= g(\chi,\psi)
\end{equation}
To quantize we complexify and decompose
\begin{equation}
V_{\Bbb C}= {\Bbb C} \otimes V = V^+ \oplus V^-
\end{equation}
This decomposition (which defines a complex structure)
\cite{Ashtekar,Woodhouse,Saunders} 
is not unique.

\vskip .5cm
{\sl This non-uniqueness corresponds physically to the possibility
of particle production and is an essential part of our current
understanding of {\bf black hole evaporation} and 
{\bf  inflationary perturbations} }

At this point it may be  instructive to 
recall \cite{Wightman} why commutation 
relations of the form
\begin{equation}
\bigl[{\hat x} ^\mu,  {\hat P} _ \nu \ \bigr]= i \delta ^\mu _\nu
\end{equation}
don't apply in quantum field theory in Minkowski spacetime. 
If they did, then 
they would have, up to natural equivalence,
  to be represented in the standard  Stone-Von-Neumann
fashion on $L^2({\Bbb E} ^{3,1} )$. But then the energy ${\hat P}^0$,  
could not be bounded below. Thus $L^2( {\Bbb E} ^{3,1} )$ is
not the quantum mechanical Hilbert space. Rather, as stated above,
it is the space of positive frequency solutions
of the Klein-Gordon or Dirac equations. These are is  much more subtle objects
and certainly not uniquely defined in a curved spacetime manifold
$\{ {\cal M},g \}$,   unlike $L^2 (M, \sqrt {-g} d^4\, x)$,
the obvious generalization of $L^2({\Bbb E} ^{3,1})$,  which
is unambiguous even in a curved spacetime.

\vskip .5cm

\subsection{The Wave Function of the Universe}

In {\it Hartle and Hawking's Wave Function for The Universe}
\begin{equation}
\Psi(h_{ij}, \Sigma ) = 
\int d[g] e^{-I_{\rm euc} (g)} \,,\quad  h_{ij}= g_{ij} |_{\Sigma =\partial M}
\end{equation}
Is real valued. To get a notion of time one typically passes
to a Lorentzian  WKB approximation $S_c$ 
\begin{equation}
\Psi = A e^{i S_c} + \bar A e^{-iS_c} 
\end{equation}
but this is only a semi-classical approximation, in other words
\vskip .5cm

{\sl Time, the complex numbers and the complex structure $i_{\rm qm}$ 
 of quantum
 mechanics
{\bf emerge} only as an approximation at late times}

\vskip .5cm

\subsection{Euclidean Quantum Field Theory}
In fact in  Euclidean Quantum Field Theory
it is not sufficient just  to compute correlators.

In order to recover{\sl Quantum Mechanics}, rather than
merely
to indulge in an unphysical case  of  {\sl Statistical
Mechanics}, the correlators must exhibit  
{\it Reflection Positivity} \cite{Glimm,Frohlich}. 
This guarantees the possibility of analytically
continuing to real time.

This can be done for Riemannian backgrounds if they admit a
suitable
reflection map, for example static or time-symmetric metrics
such as  Real tunneling geometries \cite{GibbonsHartle,GibbonsPohle,
Gibbons3,Uhlmann,Gibbons4,Angelis,JaffeRitter1,JaffeRitter2,JaffeRitter3}  

However most Riemannian metrics do not admit such a reflection map.

\vskip .5cm 
{\sl Thus {\bf generically}  in such approaches one would not recover
standard complex quantum mechanics. Only for very special 
classical saddle points of the functional integral 
would a well defined  complex structure  emerge } 
\vskip .5cm

\subsection{Lorentzian Creation {\it ex nihilo}}

The next  example involves a Lorentzian Born From Nothing
Scenario  \cite{Friedman,GottLi,Gratton}. 
Essentially, one  considers de-Sitter spacetime
modded out by the antipodal map
 $dS/{\Bbb Z}_2$ (so-called elliptic interpretation).
\begin{equation}
-(X^0)^2 + (X^1)^2  + (X^2)^2   + (X^3)^2 + (X^4)^2       
 = { 3 \over \Lambda} \,\qquad {\Bbb Z}_2 : \,  X^A \equiv -X^A\,.  
\end{equation}
Now the antipodal map preserves space orientation
but reverse time orientation. But in quantum mechanics
a time reversing transformation  is represented by
an anti-unitary operator $\Theta$ and if all states are invariant
up to a factor
\begin{equation}
\Theta \Psi =\lambda \Psi
\end{equation}
then {\it only real linear combinations are allowed}.

Thus {\sl Quantum Mechanics in $dS/{\Bbb Z}_2$ is Real Quantum
Mechanics}.

This jibes with the fact that under the action of the antipodal map
is {\sl antisymplectic} on the bosonic space of solutions $V$ \begin{equation}
\omega (\cdot  ,\cdot  ) \rightarrow -\omega (\cdot, \cdot) \,.  
\end{equation}
This renders  imposing the CCR's impossible \footnote{ Bernard Kay
has implemented this argument more rigourously within  an  algebraic
framework \cite{Kay}}

Compare  regular time reversal
\begin{equation}
(p_i,q^i) \rightarrow \omega = (-p_i,q^i ) 
\Longrightarrow d p_i \wedge d q^ i \rightarrow -d p_i \wedge dq ^i
=-\omega  \end{equation}
If there is no symplectic form then the Heisenberg 
commutation relations make no sense, one cannot geometrically quantize.

This pathology arises quite generally for spacetimes which do not
admit
a {\sl Time Orientation}, i.e. a smooth choice of future
lightcone. Such spacetimes always have a double cover which is
time orientable and so may be regarded as the quotient of
a time orientable spacetime by a generalized , time orientation
reversing, antipodal map. The double cover thus realizes
various speculative ideas of the past and not so recent past
\cite{Goldhaber,Stannard,Albrow,Schulman,VickersLandsberg} of spacetimes
in which the arrow of time runs one way in one part and the 
other way in the other.

In other words quantum field theory is not defined unless one may
define an  {\sl Arrow of Time}
 \footnote{Amusingly CTC's seem to be quiet innocuous from this point 
of view. It seems that they  can  be compatible with quantum mechanics.}.

Amusingly CTC's seem to be quiet innocuous from this point of view.
It seems that they  can be compatible with quantum mechanics,
but not necessarily locality.   

\section{Topology, Time Reversal and the Arrow of Time}

An interesting question, discussed by
Chamblin and myself \cite{ChamblinGibbons2},   is whether this arrow is
intrinsically defined, or whether both possibilities are on the same footing. 

In other words, do there exist time-orientable spacetimes which have an 
{\sl intrinsic direction of time?}

The analogy here is with a quartz crystal which is either left-handed
or right handed. This is because  the point group contains no
reflections or inversions.

For a spatial manifold $\Sigma$  one asks: does $\Sigma$ 
there exist  an orientation reversing  diffeomorphism. In other words
is there a diffeomorphism taking  $\Sigma$ with one orientation
to $\Sigma$ with the opposite orientation?.
For such manifolds  a {\it Parity Map}  cannot be defined.
Such \lq\lq handed \rq \rq 
manifolds are quite common, certain Lens Spaces and 
${\Bbb C}{\Bbb   P}^2$ being examples 
\footnote{see Hartle and Witt \cite{HartleWitt}}.  
   
For spacetimes the analogous question is 
whether there exist a time reversing diffeo  $\Theta$?

We found  some rather exotic 
examples, based on higher  dimensional Taub-NUT spacetimes
for which no such diffeo $\Theta$ exists.

The question can be formulated in {\it Hamiltonian Mechanics}.
Does there exist a symplectic manifold $\{ M,\omega \}$ 
admitting no anti-symplecto-morphism, i.e. a time reversal 
map $\Theta$ such that
\begin{equation}
\Theta ^\star \omega =-\omega\,.
\end{equation}

The answer to this topological question, which should not be confused
with  asking whether any
particular  Hamiltonian, on  a symplectic manifold which {\sl does} admit a
time reversal
map, is  time-reversal invariant, i.e. whether

\begin{equation}
\Theta ^\star H = H \, ? \,,
\end{equation}
 
does not seem to be known.

\section{Unification and ${\rm Spin} (10)$.}

If the viewpoint advocated here is on the right track,
one might expect that should be signs in what 
little information we have about possible unification schemes.
A very popular one is based on the group $SO(10)$ and it 
is perhaps gratifying that it seems to fit with the
philosophy espoused here. 

In the standard electro-weak  model, the neutrinos are purely  
left-handed
and a  description of the fundamental degrees of freedom
in terms of Weyl spinors is often felt to be appropriate.
One may then argue that this  more
more convenient with the mainly
minus  signature. However nothing prevents one describing
it using Majorana notation and the mainly plus signature.
Moreover  the discovery of the non-zero neutrino masses
and the so-called see-saw mechanism make  it plausible 
that there is a right handed partner for the neutrinos
and the fact that then each family would fit into a chiral
(i.e. ${\bf 16}$) representation of ${\rm Spin} (10)$ 
makes it perhaps more attractive to describe the
fundamental fields in Majorana notation. This would tend to favour
the use of the mainly plus signature.

To see this in more detail recall \footnote{This is clear from
the periodicity modulo eight of Clifford algebras 
${\rm Cliff}(s+8,t)\equiv {\rm Cliff}(s,t) \otimes M_{16}
 ({\Bbb R})$   and the easily verified fact
that the   that ${\rm Cliff }(2,0;{\Bbb R}) \equiv M_2({\Bbb R})$.}
\begin{equation}
{\rm Cliff}( 10,0;{\Bbb R}) \equiv M_{32}({\Bbb R})\,.
\end{equation}

Let $\Gamma_a$, $ a=1,2,\dots ,10$ be a representation  of
the  generators
by real $32 \times 32$ matrices  and
\begin{equation}
\Gamma _{11} = \Gamma _1 \Gamma _2 \dots \Gamma _{10}\,,
\end{equation}   
so that \footnote{The matrices $\Gamma _a, \Gamma _{11}$ generate
the M-theory Clifford algebra ${\rm Cliff }(10,1; {\Bbb R})
\equiv M_{32} ({\Bbb R} ) \oplus M_{32} ({\Bbb R} )$.}
\begin{equation}
\Gamma ^2 _{11} =-1\,.
\end{equation}

It is customary to describe the ${\rm Spin} (10)$
model in terms of 16 left handed spacetime  Weyl fermions which are then placed
in a single complex chiral ${\bf 16}$, $ \Psi$  of ${\rm Spin} (10)$  
\begin{equation}
\Gamma _{11} \Psi = i\Psi\,, 
\end{equation}
but this is completely equivalent ,and notationally simpler
to regard the 16 spacetime Weyl  fermions as 32 spacetime Majorana fermions
and  then  to   regard $\Psi$ as  a 
32 dimensional Majorana spinor of ${\rm Spin} (10) $ subject to the
constraint  
\begin{equation}
\Gamma _{11} \Psi = \gamma _5 \Psi\,. 
\end{equation}

In more detail, we start with the 15  observed left handed  Weyl fermions  
of the electro-weak theory with their weak hypercharges $Y=
Q - t_3$, where $Q$ is the electric charge and $t_3$ the
third component  of weak iso-spin  
\begin{equation}
\pmatrix {u _L\cr d_L }\,,\, Y= {1 \over 6} \,\qquad  
\pmatrix {v _L\cr e_L }\,,Y \,  = -{ 1 \over 2}  \end{equation} \begin{equation}
u^c  _L  \,,\,  Y= -{ 2 \over 3}  \,\qquad  d^c _L \,,\, Y = { 1 \over 3}   \qquad e^c_L\,, \,, Y= 1\,.  
\end{equation}

The first row consists of 4 iso-doublets and the second row
of 7 iso-singlets. The up and down quarks $u_L$ and $d_L$  are in a ${\bf 3}$
of $SU(3)$  colour and their charge conjugates $u^c_L$, $ d^c_L$ 
are in a    $\bar {\bf 3} $
of $SU(3)$. In fact the, because  effective group is $S (U(3) \times  U(2) ) 
\equiv SU(3) \times SU(2) \times U(1) /{\Bbb Z}_3\times {\Bbb Z}_2$,
where $ {\Bbb Z}_3$ and ${\Bbb Z}_2$ are the centres of 
$SU(3)$ and $SU(2)$ respectively \cite{Hucks}. 
This is because  the electric charge assignments  are such that 
acting with  ${\Bbb Z}_3 \times  {\Bbb Z}_2 \equiv   {\Bbb Z}_6 $
can always be compensated  by a $U(1)$ rotation.

Now $S(U(3)\times U(2))$ is a subgroup of $SU(5)$ and is well known
one may fit all 15 left handed Weyl spinors in a ${\bf 5}$ and a ${\bf
10} $. However it is more elegant to adjoin the charge  conjugate
of the  right handed neutrino, $\nu ^c_L$  
to make up a complex ${\bf 10}$ of ${\rm Spin}(10)$.
In  fact  the multiplets may be organized into multiplets
of the ${\rm Spin}(6) \times {\rm Spin}(4) \equiv SU(4) \times SU(2) 
\times SU(2) $ subgroup of ${\rm Spin}(10)$
\begin{equation}
\pmatrix {u _L\cr d_L }\,,\qquad  
\pmatrix {\nu  _L\cr e_L } \,.  \end{equation} 
\begin{equation}
\pmatrix {u^c _L\cr d ^c_L }\,,\qquad  
\pmatrix {\nu ^c   _L\cr e ^c_L } \,.
\end{equation}
In this formalism  we have left-right symmetry
with the first row consisting  of  4 weak iso-doublets
and the bottom row of 4 doublets of some other, as yet unobserved
 $SU(2)$. The quarks and leptons also form two 
${\rm Spin} (6)\equiv SU(4) $ quartets.

\section{Gravitational CP violation?}

To conclude I would like to illustrate 
once more the advantages of the reality  viewpoint
by addressing a  question of some current interest which is relevant to the
present proceedings. That is  whether
CP-violating Dirac and Majorana mass terms for spin half fermions can
give rise to detectably  different behaviour as the particles
fall in a gravitational field of a rotating body, due to the 
Lense-Thirring effect \cite{Singh1,Singh2,Pal,Singh3}.   

If they could, then a violation of the
Weak Equivalence Principle in the form of the Universality of Free Fall
would be entailed, which
seems  rather unlikely. The calculations given in \cite{Singh1,Singh2}   
are rather complicated and in view of the great
importance of the issue, it seems worth while examining the question in 
a more elementary fashion.
There are also potential implications for the quantum theory of black holes.

There are two aspects  of the problem:
\begin{itemize}
\item The emission and detection  of the fermions 
by ordinary matter
\item Their propagation from source to detector through an intervening
gravitational field. 
\end{itemize}

It is the latter which I will be discuss here If the fermions are
assumed to be electrically neutral and  with vanishing
electric and magnetic dipole moments, this is a well defined problem 
in  general relativity. Clearly if the fermions are moving
in an  
electromagnetic field and they possess electrc charges and/or 
magnetic and electric dipole moments
the conclusions might be modified, 
but then the  question is no longer one of pure
gravity.

With our conventions, A system of  $k$ Majorana fermions $\psi$ 
has  Lagrangian 
 
\ben
L= \half \psi^t C \Dslash \psi - \half  \psi ^t C \bigl 
(m_1+m_2 \gamma ^5 \bigr ) \psi\,. 
\een
where $m_1$ and $m_2$ are real symmetric $k\times k$ matrices.

The kinetic term, but not the mass term,  is  invariant under   
$SO(k)$ transformations
\ben
\psi \rightarrow O \psi\,,\qquad O ^t O =1\,.
\een
Note that one may write
\ben
O= \exp{\omega_{ij}}\,,\qquad \omega_{ij}=-\omega_{jk}\,.
\een

The kinetic term, but not the mass term  is also  invariant under   
chiral rotations
\ben
\psi \rightarrow  P \psi\,,\een 
\ben
P= \exp{\nu_{ij} \gamma ^5 }\,,\qquad \nu_{ij}= \nu_{ji}
\een

Combining these two sets of transformations we see that 
the kinetic term, but not the mass term is in fact invariant
under the action of $U(k)$, i.e. under
\ben
\psi \rightarrow  S \psi\,, 
\een
\ben
S= \exp {\bigl(  \omega_{ij} + \nu_{ij} \gamma ^5 \bigr ) }\,.
\een

The $U(k)$ invariance is perhaps more obvious
in  a Weyl basis. Since
\ben
\bigl( \gamma ^5 \bigr )^2 =-1\,,
\een 
one may regard  $\gamma ^5$ as providing a complex structure
on the space of $4k$  real dimensional   Majorana spinors, 
converting it to the  $2k$ complex dimensional
space of positive chirality  Weyl spinors for which  
\ben
\gamma^5 = i\,.
\een

Clearly $S$  then becomes the   exponential of the $k\times k$ 
anti-hermitian matrix
\ben
 \omega_{ij} + i \nu_{ij}\,.
\een  
Thus
\ben
S  S^\dagger = 1\,.
\een
The mass matrix is then a complex symmetric matrix
\ben
m= m_1+i m_2 \,, 
\een
and under a $U(k)$ transformation
\ben
m \rightarrow S^t m S\,.
\een
At this point we invoke the result of Zumino \cite{Zumino}
that $S$ may chosen to render the matrix $m$ diagonal with
real non-negative  entries.

This implies that $k$  free massive Majorana (or equivalently Weyl)  fermions
$\psi ^i$ moving in a gravitational field will satisfy
$$
\Dslash \psi ^k - \mu_k \psi ^k =0,
$$
with no sum over $k$ and where the masses $\mu_k$ may be 
taken to be real and non-negative.
There are no exotic non-trivial effects moving past a spinning object due to the
Lense-Thirring effect. In particular there are no CP violating effects
and gravity alone cannot distinguish \lq Majorana\rq  from \lq 
Dirac \rq  masses.

\subsection {Behaviour in a Gravitational field}

From now on, we assume that the mass matrix $m$ is real and diagonal.
If one  iterates the Dirac equation and uses the cyclic Bianchi
identity in a curved space one  gets
\ben
-\nabla ^2 \psi + {1\over 4} R \psi +  
m^2 \psi =0\,.
\een
As is well known, there is no  \lq gyro-magnetic \rq coupling between the
spin and the Ricci or Riemann tensors  \cite{Peres}. 
To proceed, one may pass to a Liouville-Green-Wentzel-Kramers-Brilouin
approximation of the form 
\ben
\psi = \chi e^{i S }\,.
\een
One  obtains 
\ben
\bigl( i\gamma ^\mu \partial _\mu S + m \bigr) \chi=0, 
\een
and 
\ben
\partial _\mu S \nabla ^\mu \chi =0\,.
\een
The analogue of the Hamilton Jacobi equation is 
\ben
\Bigl ( g^{\mu \nu} \partial _\mu S \partial _\nu S + m^2 \Bigr ) 
\chi =0\,. 
\label{HJ}\een

Now since $m$ is diagonal with diagonal entries $\mu_i$, say, then
each eigenspinor $\chi_i $ 
propagates independently along timelike geodesics via
\ben
\mu_i   {d x ^\mu \over d \tau} = g^{\mu \nu} \partial _\mu S\,.
\een
The spinor amplitude $\chi_i $ is parallelly transported along these
geodesics.  Of course the geodesics are independent of the 
mass eigenvalue  $\mu_i $ 
and the polarization state given by $\chi_i$. 
Indeed if the fermion starts off in a given polarization state with
(with the associated mass), it remains in it. 
In other words, at the L-G-W-K-B level, the 
Weak Equivalence Principle, in the form
of the Universality of Free Fall  holds

\section{Pure States $\longrightarrow$ Mixed States?}

 The completely thermal character of Hawking radiation
({\sl at the semi-classical level}) 
and the apparent violation of Global Symmetries
if black hole decay leaves no remnants led Hawking\cite{Hawking}
to suggest that while the standard propositional structure of 
quantum mechanics, and its complex structure, should remain
in a full quantum theory of gravity, the evolution law should change.
In particular the evolution law  should allow 
pure states to evolve to mixed states.
In what follows I shall review the formalism suggested (and now
abandoned) by Hawking and then comment on its relation to the
suggestion I am making about the complex structure of quantum
mechanics.
I shall also relate this discussion to issues of reversibility and the
arrow of time.

\subsection{Density Matrices}
Are positive semi-definite Hermitean
operators acting on a  quantum mechanical Hilbert space
${\cal H}$  with unit trace
\ben
\rho = \rho ^\dagger\,,\qquad  {\rm Tr} \rho =1\,,\qquad 
 \langle \psi | \rho \psi \rangle \ge 0\,, \quad \forall \,
 |\psi \rangle \,. 
\een
 
If one diagonalizes   
\ben
\rho = \sum_n P_n |n \rangle \langle n|
\een
where $P_n \ge 0$ is the probability one is in the (normalized )
 state $|n\rangle$, and 
\ben
\sum_n P_n =1\,.
\een 
A pure state is one for which
\ben
{\rm Tr} \rho =1\,,
\een
 in which case, all but one of the
$P_n$ vanishes and one  is in that  state with certainty.  
In a general orthonormal   basis with    one writes
\ben
\rho = \rho _{mn}  |m\rangle \otimes \langle  n|  
\een
with 
\ben
\rho_{mm} =1\,,\qquad  \rho_{mn} = {\bar  \rho} _{nm}   
\een
There is a distinguished density matrix $\iota$ 
associated with complete ignorance for which 
$P_n= {1 \over N}$ where $N={\rm dim} _{\Bbb  C} {\cal H}$, 

\subsection{Gibbs Entropy}

Normalized density matrices form a    convex  cone in the space
of all Hermitean operators
and the Gibbs entropy 
\ben
S = -{\rm Tr} \rho \ln \rho = - \sum_n P_n \ln P_n 
\een
is a convex function on the cone which is largest
at the completely ignorant density matrix $\iota$.
and vanishes for any pure state.

If $N=2$ we may set
\ben
\rho= \half \bigl (x^0 {\Bbb I }_2 + x^i \sigma _i \bigr)
\een
where $\sigma_i$, $i=1,2,3$  are Pauli matrices and the cone
corresponds to the future light cone
of four dimensional Minkowski spacetime 
\ben
x^0 \ge \sqrt{x^i x^i} =r \,.
\een
The unit trace condition implies that $x^0=1$ and thus $r-\sqrt{x^ix^i } \le 1$
One finds that
\ben
S=-\ln [ ({1+r \over 2})  ^{1+r \over 2} ({1-r \over 2} )^ {1-r \over 2} ]\,.   
\een  
The entropy is maximum at the origin and goes to zero on 
the boundary
of unit ball .

\subsection{Evolution by an S-matrix}

In general we might be interested 
in situations where there is an in and an out Hilbert space.
${\cal H} ^{\rm in }$ and ${\cal H} ^{\rm out }$ respectively.
Conventionally on thinks of ${\cal H} ^{\rm in }$ and ${\cal H} ^{\rm out }$
as being isomorphic , except possibly described in a different basis
but one could envisage more general situations.
One has an associated set of states or density matrices
for ${\cal H} ^{\rm in }$ and ${\cal H} ^{\rm out }$.
The set of such  (unormalized ) mixed states we call ${\cal N}^{\rm in}$ or
${\cal N}^{\rm out}$ respectively .   

Conventionally one postulates there is a unitary map $S: {\cal H} ^{\rm in }
\rightarrow {\cal H} ^{\rm out } $ called an S-matrix  such that  

\ben
|{\rm out  }\rangle = S |{\rm in} \rangle  
\een
 
which acts by conjugation on mixed states or  density matrices
\ben
\rho ^{\rm out} = S \rho ^{\rm in} S ^\dagger\,. 
\een

\subsection{Tracing out}

A situation which often arises is when
the out Hilbert space${\cal H} ^{\rm out }$
is a tensor product 
\ben
{\cal H} ^{\rm out }={\cal H} ^{\rm out\, 1 }\otimes {\cal H} ^{\rm out \, 2 }
\een  
An initial state $|{\rm in}  \langle$
which remains pure will have an expansion
\ben
|{\rm in } \rangle = c_{m  M} |m \rangle \otimes | M\rangle  
\een  
where $|m \rangle $ is a basis for $ {\cal H} ^{\rm out\, 1}  $
and  $|M \rangle $  a basis for $ {\cal H} ^{\rm out\, 2} $.
An observable $O_1$ which acts as the identity on ${ \cal H} ^{\rm out\, 2}  $
will have an expectation value
\ben
\langle {\rm in }| O_1 | {\rm in} \rangle 
= \rho_{mn} |m \rangle \otimes \langle n | \,, 
\een  
 where
\ben
\rho_{mn} = {\bar c}_{ m M}  c_{n M } \,,
\een 
where we use the fact that
\ben
\langle n | O_1 | m \rangle = {\rm Tr}  \bigl (   0_1 |m \rangle \otimes
 \langle n | \bigr ) \,.   
\een

In other words observations made only in $ {\cal H} ^{\rm out\, 1}  $
can tell us nothing about $ {\cal H} ^{\rm out\, 2} $
and hence will in general behave as is the final state were
mixed.   

\subsection{Evolution by an \$ matrix} 

Taking $ {\cal H} ^{\rm out\, 1}  $ to be states at infinity
and $ {\cal H} ^{\rm out\, 2}  $ horizon states
shows that in general outgoing radiation
from a black holes with a permanent horizon will be in a mixed state.

However back reaction means that the horizon is not permanent
and  the issue arises  whether taking back reaction
into account would give  a pure or a mixed state.  
 
More generally,  one may try to construct a generalization of standard
quantum mechanics in which in general  pure  states evolve 
to mixed states. One postulates that there is a linear map
$ \$  : {\cal N} ^{\rm in} \rightarrow {\cal N} ^{\rm in}$ such that
\ben
\rho^{\rm out} = \$  \rho^{\rm in} \,.
\een
One further postulates that $\$$ is hermiticity,and  trace-preserving
\bea
&(a) \qquad & (\$ \rho) ^\dagger  = \rho ^\dagger\,, \\ 
&(b) \qquad & {\rm Tr} \$\rho = {\rm Tr} \rho\,, \\
&(c) \qquad & \$ \iota = \iota \,.
\eea

One also demands  that $\$ $ takes positive semi-definite 
operators to positive semi-definite operators.

Some comments are in order.
\begin{itemize}
\item
The assumption of linearity,  is a form of 
locality assumption since it amounts to 
assuming \lq non-interference of probabilities \rq .
It should be possible to lump together 
results of two independent experiments
and obtain the same probabilities.

Thus if in one ensemble consisting of 100  states with  30   in state
1 and 70   in state 2   these go to states 3 and 4 in 45 and
55 times respectively, and in a second run of the same experiment
30   in state 1 and 70   in state 2 go to 72 and 28 in states 3 and 4 
respectively than it should be the case, if the usual idea
of probabilities is to make sense, that run in which 85=30+55 
are in state 1 and 115=70+45  in state 2, then 117=45+72
should land up in state 3 and 83=55+28 should land up in state 4.

Of course strictly speaking, 
this argument only works for {\sl commuting} density  matrices
but, by continuity  it seems reasonable to assume  linearity for  all 
density matrices.    
\item
The assumption that the  
completely ignorant density matrix $\iota$  
is preserved in time would seem to be necessary for any type
of thermodynamics  to be possible, not least because the
completely ignorant density matrix $\iota$ has the largest Gibbs entropy.  
\subsection{Invertibility and Factorisability}

Standard S-matrix evolution is such that
\ben
\$\rho = S \rho S^\dagger \,.
\een
Such $\$$-matrices are said to be {\it factorisable}
and factorisable density matrices clearly take
pure states to pure  states, but in
a general $\$$ matrix will take pure states to mixed states.
In fact, in general, a \$ matrix acts as a contraction on the
convex cone of positive definite Hermitian 
operators.Thus in general it is not invertible \cite{Wald,Page}. Indeed there is a 

\noindent {\bf Theorem}
{\it A super-scattering matrix \$ is invertible iff
it is factorisable}

\noindent{\bf Proof} Assume the contrary.  Then there exits
a mixed out-state $\rho ^{\rm out} $ 
 which is mapped to a pure state $ \$ \rho ^{\rm out} = | {\rm in}  \rangle $.
Thus \ben
\$ \sum_n P_n | n, {\rm out}  \rangle \otimes  \langle n, {\rm out} | 
= | {\rm in}  \rangle \otimes  \langle {\rm in} |\een 
 
Let $| \psi  ^{\rm in} \rangle $ be any in  state 
orthogonal to   $| {\rm in } \rangle $
One has 
\ben
\sum_n P_n \langle \psi ^{rm in} | 
\Bigr ( \$ | n,{\rm out} \rangle \langle n, { \rm out }  |
\Bigl ) | \psi ^{\rm in} \rangle =0\,.  
\een
But $  \$ | n,{\rm out} \rangle \langle n,{ \rm out }  |$
is a density matrix and so positive semi-definite. Thus  a
if $| n,{\rm out} \rangle$ has $P_n \ne 0$, then it must be  orthogonal
to every pure  state $|\psi ^{\rm in } \rangle $   orthogonal to 
$ |{\rm in}  \rangle $ and hence 
\ben
\$| n,{\rm out} \rangle = |{\rm in} \rangle 
\otimes  \langle {\rm in} |\,,\qquad \forall
\, \{ P_n | P_n \ne 0 \}\,.\een
\end{itemize}
But if $\$$ takes {\sl all}  such states $|n, {\rm out } > $ to the same 
state $| {\rm in} \rangle $ it cannot be invertible.

\subsection{Irreversibility and CPT}

Thus, as one might have expected, 
evolution by a superscattering matrix would  irreversible.
How does this square with our prejudices about $CPT$?
This is usually taken to be an anti-unitary invertible 
(since $ \theta ^2 = 1$ )   
$\theta: {\cal N} ^{\rm out  } \rightarrow 
{\cal N} ^{\rm out }$ which takes pure states to pure states,
and  preserves
traces and preserves ignorance. In fact one usually has

Let us call its restriction to pure states

\subsection{Strong CPT} assumes an invertible map
$\Theta$  from in states to out states 
\ben
\Theta = \$ \Theta ^{-1} \$\,.
\een
Thus
\ben
\$^{1} = \Theta ^{-1} \$ \Theta ^{-1} \,.
\een
In other  words Strong CPT implies that the evolution is invertible.
Note that this rather strong result does not assume
that either $\Theta$ or $\$$ is a linear map.
However if $\$$ satisfies the requirements for a 
superscattering matrix and strong CPT , then it must be  
invertible and hence factorisable.

\subsection{Weak CTP}

Faced with the result above, one could argue that only {\sl
probabilities} are related by CPT.
this
\ben
{\rm Prob} (|\psi\rangle  \rightarrow |\phi \rangle )
= {\rm Prob} (\Theta ^{-1} |
\phi\rangle  \rightarrow \Theta |\psi \rangle )\,.
\een
That is
\ben
\langle \phi |  \$ \Bigl 
( |\psi \rangle \langle \psi | \Bigr )  \phi \rangle
= \langle \Theta \phi | \$ \Bigl ( | \Theta ^{-1} \phi \rangle \langle
\Theta^{-1}  \phi | \Bigl | \Theta \psi \rangle \,, 
\een
that is
\ben
\$ ^\dagger = \Theta ^{-1} \$ \dagger \Theta ^{-1} \,.\label{weak}
\een
Of course for a factorisable $\$ $ matrix (\ref{weak}) 
holds by unitarity of the $S$ matrix. .

Moreover (\ref{weak}) implies that the superscattering
operator is ignorance preserving
\ben
\$ i =i\,.
\een

An interesting set of questions is
\begin{itemize}
\item Is (\ref{weak}) equivalent to {\it detailed balance}?
\item Does (\ref{weak}) imply the H theorem ?
\item Does (\ref{weak}) imply that only the microcanonical ensemble,
i.e. the perfectly ignorant density matrix $i$ 
is left-invariant by $\$  $?
\end{itemize}  

A full answer to these questions appears not be known but
what is well known is the situation when all density
matrices are assumed diagonal.

\subsection{Pauli Master Equation}

This is essentially the case when  the density matrix remains
diagonal. One sets
\ben
{\dot P}_r = \sum_{ s \ne r} U_{rs} P_s - P_r \sum_{s \ne r} U_{sr} \een
where $U_{rs} \ge 0 $ may be interpreted as  the transition probability per unit time
if a transition from   state $|s \rangle \langle s|$ to state
  $|r \rangle \langle r|$.

In perturbation theory
\ben
U_{rs}= | \langle r | H_{\rm pert} | s \rangle |^2
=   \langle r | H_{\rm pert} | s \rangle ^\star 
\langle r | H_{\rm pert} | s \rangle
\een
 and hence from the  Hermiticity of the Hamiltonian
\ben
\langle r |H_{\rm pert} | s \rangle ^\star = 
\langle s | H_{\rm pert} | r \rangle
\een
we have{\it detailed balance} or {\it microscopic
reversibility}
\ben
U_{rs}=U_{sr}
\een
Under this assumption and that all transitions take place, 
i.e $U_{rs} >0 \quad \forall\,  r,s $
we have 
the two following \cite{Thomsen,Aharony,AharonyNeeman1,AharonyNeeman2,Neeman}
\medskip 

\noindent{\bf Theorem A }{\it  (Existence Uniqueness of Equilibrium) 
there is a unique equilibrium state $\iota$ of total ignorance such that
$P_r=P_s\,, \quad \forall r,s $ } 
and 
\medskip

\noindent{\bf Theorem B}{\it  (\lq H-Theorem \rq ) 
The entropy $
S = - \sum_r P_r  \ln P_r 
$  is monotonic increasing 
$\dot S \ge 0\,.$}
\medskip

\noindent{\bf Proof of A } under these assumptions
\ben
\dot P_r= \sum_{s , s \ne r} U_{rs} \bigl(P_s-P_r  \bigr )  \,.
\een
If we order  the $P_s$ in numerical order  the r.h.s is non-negative
and vanishes iff $P=P_s\, \forall\, r,s$
 
\medskip

\noindent{\bf Proof of B } under these assumptions
it is also true that 
\bea
-\dot S & =& \sum_{r, s ; r \ne s} U_{rs} \bigl(P_s-P_r  \bigr ) \ln P_r   \,.
\\
&=& - \half \sum_{r,s ; r\ne s} U_{rs} (P_-P_r) \bigl( \ln P_s - \ln P_r 
\bigr) 
\eea
But
\ben
(x-y) \bigl(\ln x-\ln y \bigr ) \ge 0\,.
\een

The problem is that in general $U_{rs} \ne U_{sr}$.
In fact 
\ben
U_{rs}= | \langle r |T | s \rangle |^2\,,
\een
where the S-matrix is given by
\ben
S=1+i T \,.
\een
{\sl Unitarity}  them implies that 
\ben
\sum_s U_{rs} = \sum _r U_{rs}  \,.
\een

\subsection{Consequence of Symmetries}
It is well known that in standard $S$ -matrix  quantum mechanics that
that symmetries and conservation laws are closely related.
In the case of $\$$-matrix quantum mechanics the connection is much less close.
\subsection{$S$-matrix case}

Wigner's theorem tells us that if  $T$ be a norm preserving map 
acting on the pure  states preserving probabilities, 
then $T$ must be unitary or anti unitary $T^{-1} = T^\dagger$ .
We also assume a similar map $T^\prime$ acts   on the out pure  states
${T^\prime}  ^{-1} =  {T^\prime}^\dagger$, then if the 
$S$ matrix is invariant
\ben
S T = T^{\prime} S\,.
\een
Thus
\ben
ST S^{-1} = T^{\prime} \,.
\een
Now if
\ben
T= \exp {i\epsilon G}\,,\qquad G= g^\dagger\,,
\een
then 
\ben
S G S^{-1} = G^\prime\,,
\een
where $T^\prime = \exp {i\epsilon G ^\prime }$ .

Thus if $|{\rm out}  \rangle = \$ |{\rm in} \rangle$
\ben
\langle {\rm in} | G | {\rm in} \rangle = 
\langle {\rm out}| G^\prime |{\rm out}\rangle\,. 
\een
In other words, $H$ is conserved. More over it also
follows that any power  $G^k$ of $G$ is conserved
and that eigenstates of $H$ and  are taken to eigenstates
of $G^\prime$.    

\subsection{$\$$-matrix  case}

In the $S$-matrix case, a density matrix transforms under $T$ as
\ben
\rho \rightarrow {\cal T} \rho =T \rho T^\dagger\,.
\een
with 
\ben
{\cal T}= {\cal T} ^\dagger \,.
\een

The condition of symmetry is now
\ben
\$= {{\cal T} ^\prime}^{-1} \${ \cal T} = 
 {{\cal T} ^\prime}^{\dagger} \${ \cal T} \,.   
\een

It is easy to see with particular examples that, in general symmetries,
 do not imply
conservation laws \cite{Gross}.
\section{Superscattering and the Reals}
Hawking's original proposal( now famously abandoned by him)
 assumed the standard complex structure 
of quantum mechanics. From the point of view of
what I have been advocating it seems curiously
conservative to maintain that while 
advocating a much more radical modification
of what we mean by the laws of physics. 

In fact the entire discussion above works just as well
over the reals, that is when the density matrices are just real
symmetric semi-definite. 

The general theory of  super-scattering matrices works
over all three fields, ${\Bbb R}, {\Bbb C}$ and ${\Bbb H}$ 
and interestingly the space of such matrices
is itself a convex set. Now any convex set is, by a Theorem
of Minkowski, the convex hull of its extreme points.
In this case,  the  extreme points are unitary or anti-unitary
purity preserving maps, i.e. $S$-matrices.     
 
A simpler case to consider is restrict attention to the case of
diagonal density matrices.
In this case, $\$$ matrices are the doubly stochastic matrices encountered
in the theory of Markov processes. These are the convex hull
of the permutation matrices which take pure states to pure states.

The general theory of $\$$ matrices, at least in finite dimension,
is nicely discussed in \cite{AlbertiUhlmann}.

\section{Conclusion}

We have seen in this talk that
\begin{itemize}
\item Time and its arrow are intimately linked
with the complex nature of quantum mechanics. 
\item It is not difficult to construct spacetimes 
for which no arrow of time exists and on which 
backgrounds only
real quantum mechanics is possible
\item Only  Riemannian manifolds  admitting a reflection map $\Theta$ 
allow the recovery of standard  quantum mechanics 
\item Even if one can define an  arrow of time
it may not be possible to define an operator $\Theta$ which reverses it. 

\end{itemize}

Why then do we have such a strong  impression that time exists and
that it  has an arrow? When and how did the complex numbers get into
quantum mechanics? 

Like so many things in life: its all a matter of history.
The universe \lq\lq started \rq \rq with very special initial conditions
\lq \lq when \lq \lq neither time nor quantum mechanics
were present. Both are emergent phenomena. Both are 
consequences of the special state we find ourselves in.

Constructing and understanding that state, and its alternatives is the
on-going challenge of Quantum Cosmology.
 
\section{Acknowledgements}
I would like to thank, Thibault Damour, Stanley Deser, Marc Henneaux
and John Taylor for helpful discussions and suggestions about the material
in section 11.

\section{Appendix: Complex versus Real  Vector spaces}
In this appendix we review some mathematical facts about
complex structures.  The standard structure of quantum mechanics
requires that (pure) states are rays in a Hilbert space
${\cal H}_{\rm qm}$ which is a vector space over the complex numbers    
carrying a Hermitian positive definite  inner product $h(U,V)$ such that

\begin{eqnarray}  &(i)& \qquad  h(U,\lambda V) =\lambda h(U,V) \,,\qquad 
\forall \lambda \in {\Bbb C} \,.\\   
      &(ii)& \qquad  h(U, V) ={\overline  h(V,U)} .\\
   &(iii)& \qquad  h(U, U) >0 .\end{eqnarray}

It follows  that $h(U,V)$  is antilinear in the first slot

\begin{equation}  h(\lambda U, V) ={\overline \lambda}  h(U,V) \,,\qquad 
\forall \lambda \in {\Bbb C} \,.\end{equation} 
 
In Dirac's bra and ket notation elements of  ${\cal H}_{\rm qm}$
are written as kets:
\begin{equation}
V \leftrightarrow |V \rangle 
\end{equation}
and elements of the ${\Bbb C}$- dual space ${\cal H}_{\rm qm}^\star $,
the space of $\Bbb C$- linear maps
 ${\cal H}_{\rm qm} \rightarrow {\Bbb C}  $   as bras: 
and there is an anti-linear  map from  ${\cal H}_{\rm qm} $  to  
${\cal H}_{\rm qm}^\star $ given by
\begin{equation}
 U \rightarrow \langle U|       
\end{equation}
 such that
\begin{equation}
h(U,V) = \langle U\,|\, V \rangle\,,
\end{equation}
thus
\begin{equation}
\langle U | = h(U,\, \cdot \, )\,. 
\end{equation}
In components 
\begin{equation}
| V \rangle = V^i \,| i \rangle 
\end{equation} and 
\begin{equation}
\langle U |= \langle j |\,{\bar    U}^{\bar   j}  ,  
\end{equation}
\begin{equation}
\langle U\, |\, V \rangle = h(U ,V) =
 h_ {{\bar  i} j} {\bar  U} ^{\bar  i} V^j,
\end{equation}
where
\begin{equation}
h_ {{\bar i} j}= \langle j\,|\, i\rangle \,,
\end{equation}
and
\begin{equation}
{\overline h_ {{\bar  i} j}} = h_ {{\bar  j} i}
\,.\end{equation}

\subsection{Complex  Vector spaces as Real Vector spaces } 
A useful references for this material 
with a view to applications in
physics are  \cite{Flaherty,Trautman}. 
 
For simplicity of exposition one may imagine that ${\cal H}_{\rm qm}$
as finite dimensional 
${\rm dim}_{\Bbb C} {\cal H}_{\rm qm} =n < \infty $.   
Since a complex number is just a pair of real numbers \cite{Hamilton},
any Hermitian  vector space may be regarded as a real vector space $V$  
of twice the dimension  ${\rm dim}_{\Bbb R}\,  V  =2n $      with
something added \cite{Stueckelberg}, 
a complex structure $J$, i.e a real-linear map 
such that 
\begin{equation}
J^2 =-1\,,
\end{equation}
 and a positive definite metric .  
$g$ such that $J$ is an isometry, i.e. 
\begin{equation}
g(JX,JY) = g(X,Y) \,.
\end{equation}
It follows that $V$ is also a symplectic  vector space, with
symplectic form  
\begin{equation}
\omega(X,Y) = g(JX,Y)= -\omega (Y,X)\,,
\end{equation}
and $J$ acts canonically, i.e.
\begin{equation}
\omega(JX, JY) = \omega(X,Y)\,.
\end{equation}
Alternatively given $J$ and the symplectic form $\omega$ we obtain
the metric $g$ via
\begin{equation}
g(X,Y)=\omega (X, Jy)\,.
\end{equation}

The standard  example is the complex plane ${\Bbb C}= {\Bbb R} ^2 $ 
where if
\begin{equation}
{\bf e}_1=(1,0)\,,\qquad {\bf e}_2=(0,1)\,,
\end{equation}
\begin{equation}
J ({\bf e}_1 )=  {\bf e}_2\,,\qquad J ({\bf e}_2 )= - {\bf e}_1
\end{equation}
or as a matrix 
\begin{equation}
J = \pmatrix {0 & -1,\cr 1 & 0 \cr }
\end{equation}
and thus
\begin{equation}
J( x  {\bf e}_1 + y { \bf e}_2 ) =   x  {\bf e}_2 - y { \bf e}_1 \,
\end{equation}
which is the same in the usual notation as  
\begin{equation}
i(x+iy) = -y+ix\,,
\end{equation}
where $1 \leftrightarrow (1,0)$
and  $i \leftrightarrow (0,1)$ .

A complex structure $J$ can be thought of as a rotation
of ninety degrees in $n$ orthogonal two planes.
To specify it therefore it suffices to specify the (unordered)  
set of planes and the {\it sense} of rotation in each 2-plane. 

\subsection{A Real vector space as a Complex Vector space}

Given the original real vector space, 
how are the complex numbers actually introduced?
We start with $V$ and pass to its {\it complexification},
the tensor product  
\begin{equation}
V_{\Bbb C}= V \otimes _{\Bbb R} {\Bbb C}\,. 
\end{equation}
Note that ${\rm dim} _{\Bbb R} \, V _{\Bbb C} =4n =2 {\rm dim}
 _{\Bbb C} \, V _{\Bbb C}$,  

We now extend the action of $J$ to $ V_{\Bbb C}$, so
it commutes with $i \in {\Bbb C} $:
\begin{equation}
J \alpha X = \alpha J X\,, 
\qquad \forall \quad  \alpha \in {\Bbb C}\,,X \in {\Bbb C}\,. 
\end{equation}

We may now diagonalize $J$ over $\Bbb C$ and write
\begin{equation}
V_{\Bbb C}= W \oplus {\overline W}\,
\end{equation} 
where 
\begin{equation}
J W = i W\,,\qquad J {\overline W} =-i {\overline W}\,.
\end{equation}
Clearly $\dim _{\Bbb R}\, W = 2 n = 2 \dim _{\Bbb C}\,  W  =
 \dim _{\Bbb R}\,  V $,
and $W$ may be thought of as $V$ in complex notation.

Thus if $ X \in V$, we have that
\begin{equation}
X= {1 \over 2}(1-iJ)X  + {1 \over 2} (1+iJ) X,
\end{equation}
with $ {1 \over 2}(1-iJ)X     \in W$ and $ {1 \over 2}(1-iJ)X \in {\overline W}$ .
Vectors in $W$ are referred to as type $(1,0)$ or holomorphic
and vectors in $\overline W$ as type $(0,1)$ or anti-holomorphic.

\subsection{The metric on $V_{\Bbb C}$}
 
If $V$ admits a metric for which $J$ acts by isometries,
we may  extend the metric $g$ to all of $V_{\Bbb C}= W \oplus {\overline W} $
by linearity over $\Bbb C$ , we find that
\begin{eqnarray}
&(i)& g({\bar U}, V) = {\overline g(U, V)}\,\\ 
&(ii)& g ( U, {\bar U}) >0, \\
&(iii)& g(U,V) =0\,,  \forall U,V \in W \,,  {\rm and }\,,  \forall U,V 
\in {\overline W} \,.
\end{eqnarray}

\subsection{Negative Probabilities?}
The metric $g$  is usually assumed to  positive definite because of the demand 
that probabilities be positive and lie in the interval $[0,1]$.
This requirement has been brought into
question, notably by Feynman \cite{Feynman}. In the  context 
of vacuum energy  one should perhaps not be too quick in rejecting this
possibility since the expectation  value of the energy 
momentum tensor for negative probability states in such 
theories can of course have the opposite sign from the usual one.
This could  have applications the cosmological constant problem.

\end{document}